\documentclass[a4paper,11pt]{article}
\pdfoutput=1 % if your are submitting a pdflatex (i.e. if you have
\usepackage{jcappub} % for details on the use of the package, please
\usepackage[T1]{fontenc} % if needed
\usepackage{amsmath, amssymb, bm, graphicx, graphics, color, mathrsfs}
\RequirePackage{graphicx}
\RequirePackage{mathptmx}      % use Times fonts if available on your TeX system
\RequirePackage{flushend}
\RequirePackage[numbers,sort&compress]{natbib}
\usepackage{caption}
\usepackage{subcaption}

\title{\boldmath Ultra-compact spherically symmetric {\it dark matter}  charged star objects }

\author[1,a]{Bart{\l}omiej Kiczek\note{bkiczek@kft.umcs.lublin.pl}}
\author[2,a]{Marek Rogatko\note{rogat@kft.umcs.lublin.pl, marek.rogatko@poczta.umcs.lublin.pl}}
\affiliation[a]{Institute of Physics \\
Maria Curie-Sk{\l}odowska University \\
20-031 Lublin, pl. Marii Curie-Sk{\l}odowskiej 1, Poland}
\abstract{
We study the properties of ultra-compact spherically symmetric {\it dark matter} sector star objects, being the solution of Einstein equations with two $U(1)$-gauge
fields. One of them is the ordinary Maxwell field, while the auxiliary gauge field pertains to the {\it hidden} sector, and mimics the properties of {\it dark matter}.
The {\it visible} and {\it hidden} sectors are coupled by a kinetic mixing term with a coupling constant $\alpha$.
We also investigate the possibility of condensation of charged scalar field around the reflecting {\it dark matter} star object. It happens that 
{\it dark matter} sector, both {\it dark matter} charge and coupling constant, cause shrinking of the star radius for which the condensation may occur.
}

\begin{document}
\maketitle
\flushbottom
%%%%%
\newcommand  {\Rbar} {{\mbox{\rm$\mbox{I}\+!\mbox{R}$}}}
\newcommand  {\Hbar} {{\mbox{\rm$\mbox{I}\!\mbox{H}$}}}
\newcommand {\Cbar}{\mathord{\setlength{\unitlength}{1em}
     \begin{picture}(0.6,0.7)(-0.1,0) \put(-0.1,0){\rm C}
        \thicklines \put(0.2,0.05){\line(0,1){0.55}}\end {picture}}}
%%%%%%%%%%%%%%%%%%%%%%%%%%%%%%%%%%%%%%%%%%%%%%%%%%%%%%%%%%%%%%%%%%%%%
% other new commands
\newcommand{\be}{\begin{equation}}
\newcommand{\ee}{\end{equation}}
\newcommand{\ben}{\begin{eqnarray}}
\newcommand{\een}{\end{eqnarray}}

\newcommand{\la}{{\lambda}}
\newcommand{\Om}{{\Omega}}
\newcommand{\ta}{{\tilde a}}
\newcommand{\bg}{{\bar g}}
\newcommand{\bh}{{\bar h}}
\newcommand{\bdel}{{\bar \delta}}
\newcommand{\si}{{\sigma}}
\newcommand{\C}{{\cal C}}
\newcommand{\D}{{\cal D}}
\newcommand{\cA}{{\cal A}}
\newcommand{\cT}{{\cal T}}
\newcommand{\cO}{{\cal O}}
\newcommand{\eeo}{\cO ({1 \over E})}
\newcommand{\G}{{\cal G}}
\newcommand{\cL}{{\cal L}}
\newcommand{\cH}{{\cal H}}
\newcommand{\cE}{{\cal E}}
\newcommand{\cF}{{\cal F}}
\newcommand{\cM}{{\cal M}}
\newcommand{\cR}{{\cal R}}
\newcommand{\cB}{{\cal B}}
\newcommand{\cJ}{{\cal J}}

\newcommand{\p}{\partial}
\newcommand{\na}{\nabla}
\newcommand{\ssum}{\sum\limits_{i = 1}^3}
\newcommand{\dssum}{\sum\limits_{i = 1}^2}
\newcommand{\tal}{{\tilde \alpha}}
\newcommand{\ints}{\int_{\Sigma} d\Sigma}
\newcommand{\LieN}{{\cal L}_{N^{i}}}
\newcommand{\Lief}{{\cal L}_{\phi^{i}}}
\newcommand{\Liet}{{\cal L}_{t^{i}}}
\newcommand{\LieM}{{\cal L}_{M^{\mu}}}
\newcommand{\Lie}{{\cal L}}

\newcommand{\tj}{\tilde j}

\newcommand{\tpe}{{\tilde p}}
\newcommand{\tp}{{\tilde \phi}}
\newcommand{\tPhi}{\tilde \Phi}
\newcommand{\tpsi}{\tilde \psi}
\newcommand{\tchi}{\tilde \chi}
\newcommand{\tim}{{\tilde \mu}}
\newcommand{\tom}{{\tilde \omega}}
\newcommand{\tr}{{\tilde \rho}}
\newcommand{\tV}{{\tilde V}}
\newcommand{\tir}{{\tilde r}}
\newcommand{\rp}{r_{+}}
\newcommand{\hr}{{\hat r}}
\newcommand{\rv}{{r_{v}}}
\newcommand{\dr}{{d \over d \hr}}
\newcommand{\dR}{{d \over d R}}

\newcommand{\hhf}{{\hat \phi}}
\newcommand{\hhM}{{\hat M}}
\newcommand{\hhQ}{{\hat Q}}
\newcommand{\hht}{{\hat t}}
\newcommand{\hhr}{{\hat r}}
\newcommand{\hhS}{{\hat \Sigma}}
\newcommand{\hhD}{{\hat \Delta}}
\newcommand{\hhm}{{\hat \mu}}
\newcommand{\hro}{{\hat \rho}}
\newcommand{\hhz}{{\hat z}}

\newcommand{\hI}{\hat I}
\newcommand{\hg}{\hat g}
\newcommand{\hR}{\hat R}
\newcommand{\hD}{\hat D}
\newcommand{\hna}{\hat \nabla}

\newcommand{\tF}{\tilde F}
\newcommand{\tT}{\tilde T}
\newcommand{\tL}{\tilde L}
\newcommand{\hC}{\hat C}

\newcommand{\ep}{\epsilon}
\newcommand{\tep}{\tilde \epsilon}
\newcommand{\bep}{\bar \epsilon}
\newcommand{\ppp}{\varphi}
\newcommand{\Ga}{\Gamma}
\newcommand{\ga}{\gamma}
\newcommand{\hth}{\hat \theta}
\newcommand{\zpsi}{\psi^{\ast}}

\newcommand{\Dsl}{{\slash \negthinspace \negthinspace \negthinspace \negthinspace  D}}
\newcommand{\tD}{{\tilde D}}
\newcommand{\tB}{{\tilde B}}
\newcommand{\talpha}{{\tilde \alpha}}
\newcommand{\tbeta}{{\tilde \beta}}
\newcommand{\hT}{\hat T}

%%%%%%%%%%%%%%%%%%%%%%%%%%%%%%%%%%%%%%%%%%%%%%%%%%%%%%%%%%%%%%%%%%%%%%%%
%%%%%%%%%%%%%%%%%%%%%%%%%%%%%%%%%%%%%%%%%%%%%%%%%%%%%%%%%%%%%%%%%%%%%%%%%%%

\section{Introduction}
\label{sec:intro}

%%%%%%%%%%%%%%%%%%%%%%%%%%%%%%%%%%%%%%%%%%%%%%%%%%%%%%%%%%%%%%%%%%%%%%%%%%%%

The pursuit for an illusive ingredient of our Universe constituting over 23 percent of its mass, non-baryonic {\it dark matter}, authorizes of the most
important topic of the astronomy and physics \cite{ber18}. 
{\it Dark matter} builds a thread-like structure
of the so-called cosmic web being a scaffolding for the ordinary matter to accumulate \cite{mas07,die12}. 
For the first time Hubble Space Telescope, revealed images of the structure in question, a giant filament of {\it dark matter}, being the part of the comic web 
extending from one of the most massive galaxy clusters MACS J071 \cite{die12}.

The contemporary understanding of the creation
of structures in the Universe, the so-called $\Lambda$CDM model,
envisages that galaxies are embedded
in extended massive halos composed of {\it dark matter}.  They are surrounded by smaller {\it dark matter} sub-halos,
which are large enough to accumulate gas and dust and form satellite galaxies, orbiting around the host ones. 
However, smaller galaxies can be circled 
by much smaller sub-halo {\it dark matter} satellites, almost invisible to telescopes \cite{sta16}. It can be concluded that in the proximity of the Milky Way we can suppose 
that such kind of structures may also exist.

On the other hand, 
{\it dark matter} interaction with the Standard Model particles
is one of the main theoretical investigations of the particle physics in the early stages of our Universe \cite{reg15,ali15}. 
Several new types of fundamental particle have been claimed 
as candidates for {\it dark matter} sector, which are 
expected to interact with nuclei in suitable detecting materials on Earth. 
It was claimed only by the DAMA collaboration \cite{ber98,ber13} that they observed modulation in the 
rate of interaction events in their detectors which might be the trace of {\it dark matter} sector. Several groups want to reproduce the DAMA results but in vain \cite{cos18}.
This situation triggers the discussion concerning
the composition, interaction with ordinary matter, the self-interaction and the possible ways
to discriminate between various models of {\it dark matter} \cite{masi2015}. 

One also tries to implement physics beyond the Standard Model for the explanation of {\it dark matter} 
non-gra\-vita\-tional interactions. This fact resurgences interests in gamma rays 
emissions coming from dwarf galaxies, possible dilaton-like coupling to photons cau\-sed 
by ultra-light {\it dark matter}, and oscillations of the fine structure constant \cite{ger15}-\cite{til15}.
The Earth experiments are also exploited for the detection of possible low-energy mass of {\it dark matter} sector,
especially in  $e^+~e^-$ colliders \cite{babar14}. BABAR detector set some energy range for {\it dark photon} production, i.e.,
$0.02 < m < 10.2 GeV$, but in vain.  No significant
signal has been observed. However the new experiments are planned to cover the energy region $15 \leq m \leq 30 MeV$. 
Recently, the revision of the constraints on {\it dark photon} with
masses below $100 MeV$ was proposed based on the observation of supernova 1987A event \cite{cha17}. Collisions among galaxy clusters can also provide new tools for testing non-gravitational forces 
acting on {\it dark matter} \cite{massey15a}.

Collapse of neutron stars 
and emergence of the first star generations can deliver some other hints for these researches in question \cite{bra14}-\cite{lop14}. The existence of {\it dark matter} 
can affect
black hole growth during the early stages of our Universe. The numerical studies of {\it dark matter} and dark energy collapse and their interactions 
with black holes and wormholes
were investigated in \cite{nak12,nak15a}.

%%%%%%%%%%%%%%%%%%%%%%%%%%%%%%%%%%%%%%%%%%%%%%%%%%%%%%%%%%%%%%%%%%%%%%%%%%%%%%%
Because of the absence of evidences for the most popular particle candidates for {\it dark matter}, it was claimed the growing sense of
'crisis' in the {\it dark matter} particle community, and necessity 
of diversifying the experimental efforts. The efforts in question should accomplish
upcoming astronomical surveys, gravitational wave observatories, to deliver
us some complimentary information about the {\it dark matter} sector \cite{ber18}.

%%%%%%%%%%%%%%%%%%%%%%%%%%%%%%%%%%%%%%%%%%%%%%%%%%%%%%%%%%%%%%%%%%%%%%%%%%%
\subsection{{\it Dark matter} model}
Having all the above arguments in mind, we shall examine the problem of the influence of {\it dark matter} on the properties of spherically symmetric static charged
star-like solution.
In our research we shall consider the model of {\it dark matter} sector in which the additional $U(1)$-gauge field is coupled to the ordinary Maxwell one. 
%%%%%%%%%%%%%%%%%%%%%%%%%%%%%%%%%%%%%%%%%
The action describing Einstein-Maxwell {\it dark matter} gravity yields \cite{vac91,ach00}
\be
S = \int d^4x \sqrt{-g}~\bigg( R - \frac{1}{4}F_{\mu\nu}~F^{\mu\nu} - \frac{1}{4}B_{\mu\nu}~B^{\mu\nu} - \frac{\alpha}{4}~F_{\mu\nu}B^{\mu\nu} \bigg),
\label{gra}
\ee
where $F_{\mu \nu} = 2 \na_{[\mu }A_{\nu]}$ is the ordinary Maxwell field while $B_{\mu \nu} = 2 \na_{[\mu }B_{\nu]}$ 
stands for the auxiliary $U(1)$-gauge field mimicking the {\it {\it dark matter} } sector coupled to Maxwell one. The coupling constant is denoted by $\alpha$.
Predicted values of $\alpha$-coupling constant, being
the kinetic mixing parameter between the two $U(1)$-gauge fields, for realistic string compactifications range between $10^{-2}$ and $10^{-16}$ \cite{abe04}-\cite{ban17}.
It turns out that
astrophysical observations of $511$ eV gamma rays
\cite{integral}, experiments detecting the electron positron excess in galaxies
\cite{atic, pamela}, as well as, possible explanation of muon anomalous magnetic moment \cite{muon}, 
strongly advocate the idea of {\it dark matter} sector coupled to the Maxwell one.
%%%%%%%%%%%%%%%%%%%%%%%%%%%%%%%%%%%%%%%%%%%%%%
Moreover, the {\it kinetic mixing term} between ordinary boson and relatively light one (the {\it dark} one) arising from $U(1)$-gauge symmetry connected with a {\it hidden sector},
may cause a low energy parity violation \cite{dav12}. 
The low energy gauge interaction in the hidden sector may manifest itself by the Higgs boson decays,
and a relatively light vector boson with mass $m\ge 10GeV$ can be produced \cite{dav13}. 

The presented model has also its justification in string/M-theory, where the mixing portal (term which couples Maxwell and the additional $U(1)$-gauge field) arises
in open string theory. Both gauge states are supported by D-branes separated in extra dimensions \cite{ach16}, in supersymmetric Type I, Type II A, Type II B models,
where the massive open strings stretch between two D-branes. The massive string/brane states existence connect the different gauge sectors. 

The other cognizance of the above scenario can be achieved by 
M2-branes wrapped on surfaces which intersect two distinct codimension four orbifolds singularities. 
It happened that the natural generalization can be carried out in M, F-theory and heterotic string models.

%%%%%%%%%%%%%%%%%%%%%%%%%%%%%%%%%%%%%%%
In the considered action (\ref{gra}) the auxiliary gauge field is connected with some {\it hidden sector} \cite{ach16}.
For the first time such a model was introduced  \cite{hol86}, in order to describe the existence and subsequent integrating out of heavy bi-fundamental 
fields charged under the $U(1)$-gauge groups.
In general, such kind of terms emerge in the theories having in addition to some {\it visible} gauge group, the other one in
 the {\it hidden sector}. Such scenario is realized e.g., in compactified string or M-theory solutions generically possess
{|it hidden sectors}  which contain the gauge
fields and gauginos, due to the various group factors included in the gauge group symmetry of the {\it hidden sector}.

The {\it hidden sector} in the low-energy effective theory contains states which are uncharged under the the Standard 
Model gauge symmetry groups. They are charged under their own groups and interact with the {\it visible} ones via gravitational interaction. 
We can also think out other portals to our {\it visible sector} \cite{portal1, portal2}.
For the consistency and supersymmetry breaking \cite{abe08},
the realistic embeddings of the Standard Model in $E8 \times E8$ string theory, as well as, in type I, IIA, or IIB open string theory with branes, require the existence 
of the {\it hidden sectors}.

The organization of the paper is as follows. In section 1 we give some introduction to the presented model of {\it dark matter} sector.
In section 2 we present the direct derivation of static spherically symmetric solution of Einstein - {\it dark matter} equations of motion.
Section 3 is devoted to the analysis of {\it dark matter} charged fluid sphere, while in section 4 we derive the bound on
the gravitational mass and charges for a stable {\it dark matter} compact star-like object. In section 5 we considered the problem of hair growing of
charged under {\it visible} sector scalar field, on the ultra-compact reflecting star. In section 6 we concluded our investigations.

%%%%%%%%%%%%%%%%%%%%%%%%%%%%%%%%%%%%%%%%%%%%%%%%%%%%%%%%%%%%%%%%%%%%%%%%%
\section{Ultra-compact charged star-like object in {\it hidden sector}}

In this section we consider the spherically symmetric static solution of Einstein {\it dark matter} sector equations of motion, 
being a model of ultra-compact charged {\it dark matter} star-like object.

The components of energy momentum tensor defined as $T_{\mu \nu} = - \delta S/\sqrt{-g} \delta g^{\mu \nu}$, 
devoted to Maxwell, {\it dark matter} $U(1)$-gauge field and the mixture of them,
may be written as
\be
T_{\mu \nu} = T_{\mu \nu} (F) + T_{\mu \nu} (B) + \alpha T_{\mu \nu} (F,~B),
\ee
where the adequate energy momentum tensors are given by
\ben
T_{\mu \nu} (F) &=& \frac{1}{2} F_{\mu \ga} F_\nu{}^{\ga} - \frac{1}{8} g_{\mu \nu} F_{\alpha \beta} F^{\alpha \beta},\\
T_{\mu \nu} (B) &=& \frac{1}{2} B_{\mu \ga} B_\nu{}^{\ga} - \frac{1}{8} g_{\mu \nu} B_{\alpha \beta} B^{\alpha \beta},\\
T_{\mu \nu} (F,~B) &=& \frac{1}{2} F_{\mu \ga} B_\nu{}^{\ga} - \frac{1}{8} g_{\mu \nu} F_{\alpha \beta} B^{\alpha \beta}.
\een
In what follows we suppose that the existence of $t$-component of the gauge field potentials
\be
A_t = - \frac{Q}{r}, \qquad B_t = - \frac{Q_d}{r},
\label{pot}
\ee
where $Q$ is the ordinary Maxwell charge while $Q_d$ is connected with {\it dark matter} sector.

The line element of spherically symmetric spacetime implies
\be
ds^2 = - e^{2 \phi(r)} dt^2 + e^{2 \la(r)} dr^2  + r^2 (\sin^2 \theta d \phi^2 + d \theta^2 ).
\label{met}
\ee
Having in mind (\ref{pot}) and (\ref{met}), the $(t~t)$ and $(r~r)$ components of
Einstein equations $G_{\mu \nu} = T_{\mu \nu}$, are provided by
\ben
- e^{2 \phi (r) } \Big[ \frac{1}{r^2} &-& \frac{e^{2 \la(r)}}{r^2} - \frac{2 \la'(r)}{r} \Big] = \frac{Q_c^2}{4 r^4},\\
\frac{1}{r^2} &-& \frac{e^{2 \la(r)}}{r^2} + \frac{2 \phi'(r)}{r} = - \frac{Q_c^2}{4 r^4 e^{2 \phi}},
\een
where the total charge $Q_c^2 = Q^2 + Q_d^2 + \alpha Q Q_d$, the prime denotes taking derivative with respect to r-coordinate.
Consequently, analyzing the above equations, we draw a conclusion that
 $\la(r) = - \phi(r)$, and the metric tensor component yields
\be
e^{-2 \la(r)} = 1 - \frac{2 M}{r} + \frac{Q_c^2}{4 r^2},
\label{lele}
\ee
where $M$ is the total gravitational mass and $Q_c$ the total charge, measured by the observer
at future timelike infinity.

%%%%%%%%%%%%%%%%%%%%%%%%%%%%%%%%%%%%%%%%%%%%%%%%%%%%%%%%%%%%%%%%%%%%%%%%%
\section{{\it Dark matter} charged fluid sphere}
This section will be devoted to the examination of the charged {\it dark sector} fluid sphere and the Tolmann-Openheimer-Volkov (TOV) equation
for the object under inspection. In order to derive the relativistic hydro and electro-{\it dark matter} dynamic equation for the charged sphere
let us consider the action for {\it visible} and {\it hidden} sector $U(1)$-gauge fields with current sources.
It is provide by the relation of the form as
\be
S_{EM+dm+ cur}(F,~B) = \int d^4x \sqrt{-g} \Big[ - \frac{1}{4} F_{\mu \nu}F^{\mu \nu}  - \frac{1}{4} B_{\mu \nu}B^{\mu \nu} - 
\frac{\alpha}{4} F_{\mu \nu}B^{\mu \nu} - 2 A^\mu j_{\mu}(F) - 2 B^\mu \tj_\mu (B) \Big].
\ee
Varying the underlying action with respect to the adequate gauge fields we arrive at the following equations of motion
\ben
\na_\mu F^{\mu \nu} &+& \frac{\alpha}{2} \na_\mu B^{\mu \nu} = 2 j^\nu (F),\\
\na_\mu B^{\mu \nu} &+& \frac{\alpha}{2} \na_\mu F^{\mu \nu} = 2 \tj^\nu (B).
\een
It can be shown that for the Maxwell field it implies
\be
\talpha \na_\mu F^{\mu \nu} = 2 j^\nu (F) - \alpha \tj^\nu (B),
\label{eqf}
\ee
where $\talpha = 1 - \alpha^2/4$ and on the right-hand side we get currents responsible both for visible and {\it dark matter} sector.
The adequate components of the energy momentum tensor imply
%defined as $T_{\mu \nu} = - \delta S_m/\sqrt{-g} \delta g^{\mu \nu}$, for the gauge field action  may be written as
%\be
%T_{\mu \nu} = T_{\mu \nu} (F) + T_{\mu \nu} (B) + \alpha T_{\mu \nu} (F,~B),
%\ee
%where the adequate energy momentum tensors are given by
\ben
T_{\mu \nu} (F) &=& \frac{1}{2} F_{\mu \ga} F_\nu{}^{\ga} - \frac{1}{8} g_{\mu \nu} F_{\alpha \beta} F^{\alpha \beta} - g_{\mu \nu}~ A_\delta 
j^\delta (F) + 2 A_{(\mu} j_{\nu)}(F),\\
T_{\mu \nu} (B) &=& \frac{1}{2} B_{\mu \ga} B_\nu{}^{\ga} - \frac{1}{8} g_{\mu \nu} B_{\alpha \beta} B^{\alpha \beta} -  g_{\mu \nu} ~B_\delta 
\tj^\delta (B) + 2 B_{(\mu} \tj_{\nu)}(B),  \\
T_{\mu \nu} (F,~B) &=& \frac{1}{2} F_{\mu \ga} B_\nu{}^{\ga} - \frac{1}{8} g_{\mu \nu} F_{\alpha \beta} B^{\alpha \beta}.
\een
Moreover, for the charged sphere we shall consider perfect fluid component of the energy momentum tensor. It yields
\be
T_{\mu \nu} (p,~\rho) = (\rho + p) u_\mu u_\nu + p g_{\mu \nu},
\ee
where $p(r)$ is the fluid pressure while $\rho$ stands for the total mass density of the sphere in question.

Having in mind that the considered ultra compact object has the spherical symmetry, the elaborated electric Maxwell field ought to posses the radial symmetry.
Just the electromagnetic strength tensor will be of the form as $F_{rt} = E(r)$. Then using equation (\ref{eqf}) we get the following relation:
\be
E(r) = \frac{1}{r^2} e^{\phi(r) + \la(r)} Q_c(F,~B)(r),
\ee
where $Q_c(F,~B)(r)$ yields
\be
Q_c(F,~B)(\zeta) = \int d\zeta~\zeta^2 ~e^{\phi(\zeta) + \la(\zeta)} j^t (F,~B)(\zeta),
\ee
while the currents for the mixture of Maxwell and {\it dark matter} fields is written in the form as follows:
\be
j^t (F,~B) = \alpha_1 j^t(F) - \alpha_2 \tj^t(B),
\ee
with the adequate coefficients $\alpha_1 = 2/\talpha$ and $\alpha_2 = \alpha/\talpha$.

Using the Einstein equation $(t t)$-component provided by the equation
\be
- \frac{e^{-2 \la(r)}}{r^2} + \frac{1}{r^2} + \frac{2 \la(r)' e^{-2 \la(r)}}{r} = \rho + \frac{Q_c^2(r)}{4 r^4},
\ee
we arrive at the final expression for the metric tensor component describing {\it dark sector} ultra compact charged sphere of a radius $r$
\be
e^{-2 \la(r)} = 1 - \frac{2 m(r)}{r} - \frac{\cF_c(r)}{r},
\label{fc}
\ee
where we set 
\be
m(\zeta) = \int d\zeta ~\frac{\rho~\zeta^2}{2}, \qquad \cF_c(\zeta) = \int d\zeta ~\frac{Q_c(\zeta)}{4 \zeta^2}.
\ee
$m(r)$ stands for the interior mass accumulated in a sphere of r-radius. It will be also necessary to introduce the notion of gravitational mass.
To commence with, let us match the exterior and interior solutions of the considered field equations. Namely at the distance $R=r$, we have equality
\be
1 - \frac{2M}{R} + \frac{Q_c^2}{4R^2} = 1 - \frac{1}{R}~\int_0^R d\zeta \rho \zeta^2 - \frac{1}{R} ~\int_0^R d \zeta \frac{Q_c(\zeta)^2}{4 \zeta^2},
\ee
which provides the explicit relation for gravitational mass $M$ at the distance equal to $R$
\be
M = \frac{Q_c^2}{8 R} + m(R) + \frac{\cF_c(R)}{2}.
\ee
Just the definition of gravitational mass at any radius $r$ implies
\be
m_g(r) = m(r) + \frac{\cF_c(r)}{2} + \frac{Q_c(r)^2}{8r}.
\ee
All the above lead to the conclusion that the gravitational mass will modify the metric function $e^{-2 \la(r)}$. Now, consequently with the help of the aforementioned notion,
we obtain that the following relation:
\be
e^{- 2 \la(r)} = 1 - \frac{2 m_g(r)}{r} + \frac{Q_c(r)^2}{4 r^2},
\ee
where the metric component depends on the gravitational mass and the total  (Maxwell and {\it dark matter} charges) of the examined sphere.

%%%%%%%%%%%%%%%%%%%%%%%%%%%%%%%%%%%%%%%%%%%%%%%%%%%%%%%%%%%%%%%%%%%%%%%%%%%%%%%%%
\subsection{TOV equation}
In order to find the Tolmann-Oppenheimer-Volkov equation for the {\it dark sector} ultra-compact charged star
we have to use the conservation of the energy momentum tensor, $\na_{\mu} T^{\mu \nu} = 0$.
Namely having in mind this relation
\be
\na_r \Big[ T_r{}^r(F,~B) + T_r{}^r\rho,~p)\Big] = 
- \frac{ Q_c(r)' Q_c(r)}{2 r^4} + p(r)' + \phi(r)' \Big(\rho + p(r)\Big)= 0.
\ee
and the $G_{rr}$ component of the Einstein equation, as well as, relation (\ref{fc}), we arrive at the following
form of the equation binding pressure, density and charge of the {\it dark sector} star-like object 
\be
p'(r) =  \frac{ Q_c(r)' Q_c(r)}{2 r^4} - \frac{\Big(\rho + p(r) \Big)}{r^2}
\Big[ m(r) + \frac{\cF_c(r)}{2} + \frac{r^3 p(r)}{2} - \frac{Q_c(r)^2}{8 r} \Big]
\Big[ 1 - \frac{2 m(r)}{r} - \frac{\cF_c(r)}{r} \Big]^{-1}.
\label{tov}
\ee

%%%%%%%%%%%%%%%%%%%%%%%%%%%%%%%%%%%%%%%%%%%%%%%%%%%%%%%%%%%%%%%%%%%%%%%%%%%%%%%
\section{Bound on the gravitational mass and charge of stable {\it dark matter} sector compact object}
In this section we look for the upper bound on gravitational mass for given radius and total charge, for the ultra-compact {\it dark matter} sector star objects.
It turns out that following the procedure presented in \cite{kar08, and09}, after some algebra, one can get almost the same result. Namely for sphere of radius $r$
the inequality implies
\be
\sqrt{m_g(r)} \le \frac{\sqrt{r}}{3} + \sqrt{\frac{r}{9} + \frac{Q_c^2(r)}{12 r}}.
\ee
On the other hand, in order to derive the inequality for the case when $r \ge R$, let us suppose that $\psi_k = m_g(r) + r^3 p(r) - Q_c^2(r)/4r$ is a sequence
of regular solutions with support in the range $[R_k,~R]$ of the Tolmann-Oppenheimer-Volkov equation (\ref{tov}). Moreover, we suppose that
$\lim_{k \rightarrow \infty} R_k/R =1$. Then, following the steps presented in \cite{and09}, one arrives at the inequality among $M,~R,~Q_c$ valid for the exterior
solution of the Einstein equations with two $U(1)$-gauge fields. Namely, it implies
\be
\sqrt{M} \le \frac{\sqrt{R}}{3} + \sqrt{\frac{R}{9} + \frac{Q_c^2}{12 R}}.
\ee
However, an interesting problem was suggested in \cite{hod18}, i.e., it was posed a question if the addition of some characteristics of horizonless object could 
improve the upper bound on its gravitational mass.
The refinement of the bounds derived in \cite{and09} was presented in \cite{hod18}.

On the other hand, geodesic motion determines important features of the spacetime and objects in it. Namely, null unstable geodesics are closely connected
with the appearance of a compact object visible by the external observer. They are also bounded with the characteristic modes of black holes, i.e.,
computing the Lyapunov exponent one can prove that, in the eikonal limit, quasi-normal modes of black holes are determined by parameters of the circular null geodesics
\cite{car09}. Circular geodesics are of special interests. For example in Kerr spacetime the binding energy of the last stable circular timelike geodesics is related
to gravitational binding energy that can be radiated to infinity. This fact can be implemented to the estimation of a spin of astrophysical black objects through
observations of the accretion discs surrounded the compact object in question \cite{nar05}.

Circular null orbits ( photonspheres) were also analyzed in the background of a regular horizonless ultra-compact stars \cite{hod18,pen19},
where it was revealed that photonspheres attributed to the aforementioned objects possessed an upper bound expressed in terms of mass and charge.

To proceed further,
let us discuss the upper bound of the gravitational mass allowed for the ultra-compact spherical {\it dark matter} sector star objects. To begin with, one finds that
the Lagrangian describing the geodesics in the spacetime of {\it dark matter} ultra-compact star yields
\be
2 \cL = - e^{2 \phi(r)} {\dot t}^2 + e^{2 \la(r)} {\dot r}^2 + r^2 {\dot \theta}^2 + r^2 \sin \theta {\dot \phi}^2,
\ee
where the dot means taking derivate with respect to the proper time. We shall restrict our considerations to the equatorial orbits. Then,
following the methods of classical mechanics, the generalized momenta with respect to the coordinates $(t,~r,~\phi)$
\be
p_t = - e^{2 \phi(r)} {\dot t} = - E, \qquad p_r = e^{2 \la(r)} {\dot r} = L, \qquad p_\phi = r^2 {\dot \phi},
\ee
where the constants $E$ and $L$ are attributed to the total energy and orbital momentum of the system, respectively. The Legendre transformation leads to 
the Hamiltonian of the form
\be
2 H = - E {\dot t} + L {\dot \phi} + e^{2 \la(r)} {\dot r}^2 = \ep = const.
\label{ham}
\ee
If $\ep =0$, one obtains null geodesics, whereas $\ep =1$ we have to do with the timelike ones.
Consequently from equation (\ref{ham}) we get
\be
({\dot r})^2 = \frac{1}{e^{2 \la(r)}} \Big[ E^2 e^{-2 \phi(r)} - \frac{L^2}{r^2} - \ep \Big].
\ee
The right-hand side of the above relation authorizes the effective potential which will be denoted by $V_r(r)$.

The condition for the circular null orbits, $V_r = V_r' = 0$, implies
\be
\frac{E^2}{L^2} = \frac{e^{- 2 \la(r)}}{r^2}, \qquad 1 + r~\la'(r) = 0,
\ee
which gives us the characteristic relation $N(r=r_c) =0$, with the definition
\be
N(r) = 3 e^{- 2 \la(r)} -1 - r^2 \Big( p(r) - \frac{Q_c(r)^2}{8 r^4} \Big).
\label{fun}
\ee
One can see that the light rings of the spherically symmetric ultra-compact {\it dark matter} sector object, which line element is provided by (\ref{met}),
are characterized by the functional (\ref{fun}). The dimensionless function $N(r)$ determines the discrete radii of the null circular geodesics of the spacetime in question.
It has the following boundary conditions
\be
N(r \rightarrow 0) = 2, \qquad N(r \rightarrow \infty) = 2.
\ee
The above feature implies that spherically regular horizonless compact objects are characterized by an even number of null circular geodesics.
Let us remark that in \cite{car14,cun17} it is derived that horizonless matter configurations having a light ring must  have pairs of them. For such regular configurations
one of the closed light ring is stable. However, there is an exception to this rule, i.e., the spherical symmetric case being subject to the equation
$\sim r_c~(\rho + p_T) =1$, where $r_c$ is the radius of the light ring and $p_T$ tangential pressure, we have odd number of the light rings 
\cite{hod18pl}.

For the exterior region of the ultra-compact object in question, one has that
\be
r \ge R, \qquad p(r) = \rho = 0.
\ee
Further suppose that the spatially regular charged {\it dark matter} sector compact object, has the external light ring for $r_c > R$.
Inspection of the equations (\ref{lele}) and (\ref{fun}) reveals that
\be
r_c(out) = \frac{6 + \sqrt{36 M^2 - 7 Q_c^2}}{4},
\ee
which gives us the stronger bound for mass and the total charge. Namely,
we arrive at the condition
\be
\frac{\sqrt{Q^2 + Q_d^2 + \alpha Q Q_d}}{M} \le \frac{6}{\sqrt{7}}.
\ee
Let us analyze the derived condition for the specific choice of the charges from {\it visible} and {\it hidden } sectors. Namely,
if $Q=Q_d$ we get
\be
\frac{Q}{M} \le \frac{6}{\sqrt{7 (2 + \alpha)}}.
\ee
The larger $\alpha$ one takes into account, the smaller value of the ratio we get.
On the other hand, when $Q_d = \beta Q$ and $\alpha = const$, we obtain
\be
\frac{Q}{M} \le \frac{6}{\sqrt{7 (1 + \beta^2 + \alpha \beta^2)}}.
\ee
Just the bigger multiplying factor $\beta$ one takes, the smaller value of $Q/M$ we have.

In principle the presence of the light ring outside the surface of a spherically regular horizonless compact star object, may cause the appearance of the another
light ring which is included inside the outer one. It was revealed \cite{kei16} that the existence of the inner stable null circular geodesic, might show that the ultra-compact object 
under consideration was nonlinearly unstable to perturbations massless  fields. Therefore we draw a conclusion that spherically symmetric horizonless spacetimes 
describing a compact object within the theory with {\it dark matter} sector has to have no light rings. It implies that the lower on the radii of the stable {\it dark matter}
star-like object yields
\be
R >  \frac{6 + \sqrt{36 M^2 - 7 Q_c^2}}{4}.
\label{rr}
\ee
Having in mind the exact form of the total charge, from (\ref{rr}), one can remark that the radius of ultra-compact {\it dark matter} sector star object will be smaller 
that this without {\it hidden sector}.

One has that the larger {\it dark charge} $Q_d$ ($M, ~Q,~\alpha = const)$ we consider, the smaller radius one obtains. Consequently,
if we assume that mass and the charges of the object are constant, and the bigger value of $\alpha$ one takes, the 
smaller radius of ultra-compact {\it dark matter} sector we receive.

%%%%%%%%%%%%%%%%%%%%%%%%%%%%%%%%%%%%%%%%%%%%%%%%%%%%%%%%%%%%%%%%%%%%%%%%%%%%
%%%%%%%%%%%%%%%%%%%%%%%%%%%%%%%%%%%%%%%%%%%%%%%%%%%%%%%%%%%%%%%%%%%%%%%%%%%%
\section{Scalar hair on {\it dark matter} compact star}
The no-hair theorem for asymptotically flat horizonless neutral compact star being subject to the reflecting boundary conditions, with the influence of
scalar field with potential, was derived in \cite{hod16}. Then the problem was elaborated in \cite{bha17}, implementing the spacetime with a positive
cosmological constant. The case of the massless scalar field non-minimally coupled to gravity was studied in \cite{hod17}.

On the other hand, the charged reflecting sphere influenced by scalar field was examined in \cite{hod16b}, while the analytical formulae for the discrete
spectrum of star radii were revealed in \cite{hod17b}. It turns out that the low frequency scalar perturbations in the spacetime of a black hole in a box can lead
to the superradiant instability, as well as, formation of a quasi-local hair \cite{san16}. It is believed that box boundary may enforce fields to bounce back and trigger the 
condensation around the black object. This kind of a problem was analyzed in the case of scalar field configurations near charged compact reflecting star \cite{pen18},
where the lower and upper bounds for the radii of the object were provided.

%%%%%%%%%%%%%%%%%%%%%%%%%%%%%%%%%%%%%%%%%%%%%%%%%%%%%%%%%%%%%%%%%%%%%%%%%%%%
\subsection{Model of scalar field configuration}
\label{sec:model}
% Action
To find out the possibility of formation of the scalar hair we start our considerations with the following action:
\begin{equation}
S _{EM+dm+sc}=  \int d^4x \sqrt{-g} \Big(-\frac{1}{4}F_{\mu\nu}F^{\mu\nu} - \frac{\alpha}{4}B_{\mu\nu}F^{\mu\nu} - \frac{1}{4}B_{\mu\nu}B^{\mu\nu} -  D_\alpha \psi 
D^\alpha \psi - \mu^2 \psi^2\Big),
\label{scf}
\end{equation}
where the covariant derivative is given by $D = \nabla_\mu - q A_\mu$. $q$ and $\mu$ are respectively charge and mass of the scalar field $\psi(r)$.
The scalar field in question is charged with respect to {\it visible } sector.

For the brevity of further notation we rename the left-hand side of the relation (\ref{lele}), i.e., $e^{-2 \la(r)} = f(r)$. Moreover, for the convenience of 
numerical calculations we also rescale 
charges in the line element describing {\it dark matter} charged static line element
\begin{equation}
\frac{Q_{(i)}}{2} \rightarrow Q_{(i)},
\end{equation}
where the subscript $(i)$ denotes charges bounded with {\it visible} and {\it hidden} sectors.

Having in mind the form of the action (\ref{scf}), we get equation of motion for the scalar field, $\delta S/\delta \psi = 0$, which is provided by
\begin{equation}
\nabla_\mu \nabla^\mu \psi - \Big( q^2 A_\mu A^\mu + \mu^2 \Big)\psi = 0.
\end{equation}
Its explicit  for the static background implies the second order ordinary differential equation of the form as follows:
\begin{equation}
\psi'' + \left(\frac{f'}{f} + \frac{2}{r}\right)\psi' + \Big(\frac{q^2 Q^2}{r^2 f^2} - \frac{\mu^2}{f}\Big)\psi = 0,
\label{eq:eom}
\end{equation}
where prime denotes derivative with respect to $r$-coordinate. It can be noticed that the relation (\ref{eq:eom}) reveals the scaling symmetry of its
parameters. Namely, one has that
\begin{equation}
r \rightarrow sr,~M \rightarrow sM,~Q_{(i)} \rightarrow sQ_{(i)},~q \rightarrow q/s,~\mu \rightarrow \mu/s.
\label{eq:scale}
\end{equation}
In order to solve the differential equation we have to establish the adequate boundary conditions imposed on $\psi(r)$. In what follows we implemet
the reflecting boundary conditions, which means that at the surface of the {\it dark matter} star the scalar field will vanish, $\psi(r_s) =0$.
Moreover, one supposes that the timelike box boundary conditions should to be satisfied in the considered spacetime. It means that at $r=r_b$ the scalar
field in question is reflected back, $\psi(r_b)=0$.

The function describing the scalar field has to have at least one extremum point in the range from $r_s$ to $r_b$. At the aforementioned point 
the function $\psi(r_{ext})$ should satisfy the following conditions:
\be
\psi' (r_{ext}) = 0, \qquad \psi(r_{ext}) \psi''(r_{ext}) \le 0.
\label{psiprime}
\ee

Following the procedure presented in \cite{pen18}, %Hod PRD i Peng
we apply the transformation $\tilde{\psi} = \sqrt{r}\psi$ in the equation \eqref{eq:eom} and
then collecting terms with $\tpsi$ derivatives, we use the extremum condition (\ref{psiprime}) in order to achieve
the inequality valid for $r = r_{extr}$
\begin{equation}
 \mu^2 r^2 f  \le -\frac{1}{4}f^2 - \frac{1}{2}f' f r + q^2 Q^2. 
\end{equation}
In the next step we check the monotonicity and the sign of each term on the right-hand side of the inequality, because of the fact that both $f(r)$ and $r f'(r)$ are descending and greater than zero, and moreover left-hand side is a descending function for the distance greater than the radius of the star. Consequently, one concludes that
\begin{equation}
\mu^2 r_s^2 f(r_s) \le \mu^2 r^2 f(r) \le q^2 Q^2.
\end{equation}
Neglecting the middle term we solve the inequality
\begin{equation}
\mu^2 r_s^2 \left(1 - \frac{2M}{r_s} + \frac{Q_{c}^2}{r_s^2}\right) \leqslant q^2 Q^2
\label{eq:upperbound_ineq}
\end{equation}
treating $\mu r_s$ as the variable in the quadratic inequality.
Finally it leads to the possible range where $r_s$ lies within. It is provided by
\begin{equation}
\mu M + \sqrt{\mu^2 (M^2 - Q_{c}^2)} \leqslant \mu r_s \leqslant \mu M + \sqrt{\mu^2 (M^2 - Q_{c}^2) + q^2 Q^2}.
\label{eq:rsrange}
\end{equation} 
The lower bound is a Schwarzschild radius for charged {\it dark matter} sector black hole and the upper one is connected with the larger solution of the inequality \eqref{eq:upperbound_ineq}.

%%%%%%%%%%%%%%%%%%%%%%%%%%%%%%%%%%%%%%%%%%%%%%%%%%%%%%%%%%%%%%%%%%%%%%%%%%%%%%%%%%%%%%%%%%%%%%%%%%%%%%%%%%%%%%%%
\subsection{Numerical results}
Integrating from $r_s$ to $r_b$,
we numerically solve the ordinary differential equation (\ref{eq:eom}) governing the motion of the scalar field $\psi$ in the spacetime of ultra-compact {\it dark matter} star-like object.
By virtue of the shooting method, we have found the adequate radius of the star surface $r_s$, with previously set $r_b$ fulfilling the auxiliary condition $\psi(r_b) = 0$ on the second boundary.
We perform our shooting by bisection method in the range derived from the maximum point analysis \eqref{eq:rsrange}.
The algorithm may find more than one value of $r_s$, corresponding to the boundary condition in a given interval. If such situation takes place, we choose the largest solution.

\begin{figure}
\centering
\includegraphics[width=.6\textwidth]{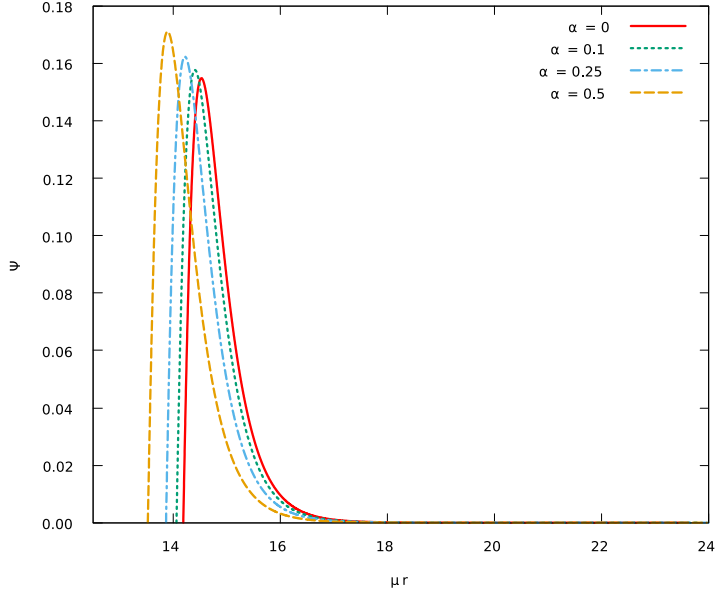}
\caption{The distribution of the condensed scalar field $\psi$ around a compact charged star-like object.
The calculation was carried out for parameters $M = 8,~Q = 4,~Q_d = 4,~q = 1$ (normalized to the mass $\mu$).
}
\label{fig:psi}
\end{figure}

%\begin{figure}

%	\begin{subfigure}{.5\textwidth}
%	\includegraphics[width=1.\textwidth]{rs_Q.eps}
%	\caption{Star radius dependence on an electric charge and the dark matter coupling constant. For this plot $M=8, q=1, Q_d = 3$.}
%	\label{fig:rs_Q}
%	\end{subfigure}
%	\begin{subfigure}{.5\textwidth}
%	\includegraphics[width=1.\textwidth]{rs_Qd.eps}
%	\caption{Star radius dependence on a dark matter charge and the coupling constant. $M=8, q=1, Q = 3$.}
%	\label{fig:rs_Qd}
%	\end{subfigure}
%\end{figure}

\begin{figure}
\centering
\includegraphics[width=1.\textwidth]{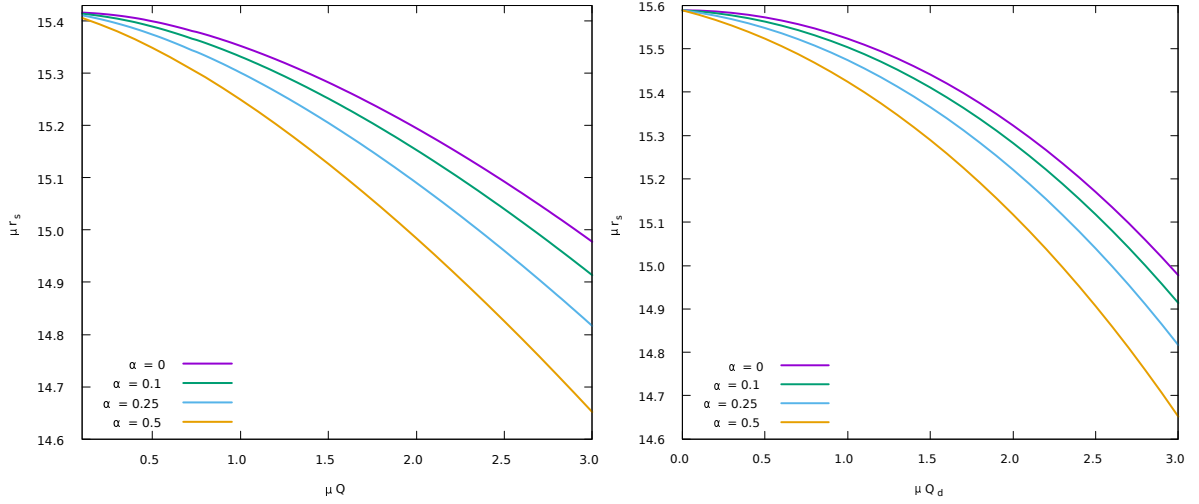}
\caption{Star radius dependence on an electric charge (left panel), {\it dark matter} charge (right panel) and the dark matter coupling constant. For this plot $M=8, ~q=1$. The unvaried charge is equal $Q_{(i)} = 3$, in both panels.}
\label{fig:rs_Q_Qd}
\end{figure}

After obtaining solution of the equation of motion satisfying the aforementioned requirements, we shall investigate 
the influence of the {\it dark matter} charge and the coupling strength on the possibility of the scalar hair formation in the vicinity
of ultra-compact star-like object. In figure \ref{fig:psi} one depicts the value of $\psi$ scalar field versus $\mu r$, for the fixed values of mass of the star and charges
of {\it visible} and {\it hidden} sectors equal to 4. The physical parameters of the system have been normalized to $\mu$ due to the scaling symmetry \eqref{eq:scale}. One can observe that the larger
$\alpha$-coupling constant we examine, the smaller radius of the hairy compact star we arrive at. The influence of {\it dark matter}
causes that the condensation takes place for the smaller objects. On the other hand, the maximal value of $\psi$ also depends on the $\alpha$-coupling constant.
If $\alpha$ increases the maximal value of the scalar field that form hair also grows.

To proceed further,
let us take a closer look on the scenario when one of the charges dominates the other. We set one of them to $Q_{(i)} = 3$ and plot the star 
radius for condensing scalar, for different values of the other kind of charge.
In figure \ref{fig:rs_Q_Qd} the dependence of the star radius on the charges, the electric and the {\it dark matter} one, is envsaged respectively.
One can see that increasing the charge forces the star to be smaller in order to hold the condensed scalar. Every curve separately shows similar 
parabolic behavior, but when we are aware of the coupling presence there are some differences to be spotted.
The effect of the coupling is similar in the change of both charges. The splitting distance between line of interest and baseline scales more or less linear with $\alpha$.
Both plots tend to the same final value of $r_s$ when the charges are equal ($Q = 3, ~Q_d = 3$) but the curves differ quantitatively. The initial value of $r_s$ from the left panel of figure \ref{fig:rs_Q_Qd} is smaller than the one from the right panel. In such a case the increase of $Q_d$ causes more steep drop of $r_s$ than in case of increasing $Q$.

The larger value of $\alpha$-coupling constant we consider, the smaller $\mu r_s$ we get.
This leads to the conclusion that the both charges result in shrinking the star radius so the scalar hair may occur.

%The scalar field feels the dark charge only through the gravitational background when the normal charge is also present in the equation of motion due to the Coulomb interaction.

\begin{figure}
\centering
\includegraphics[width=1.\textwidth]{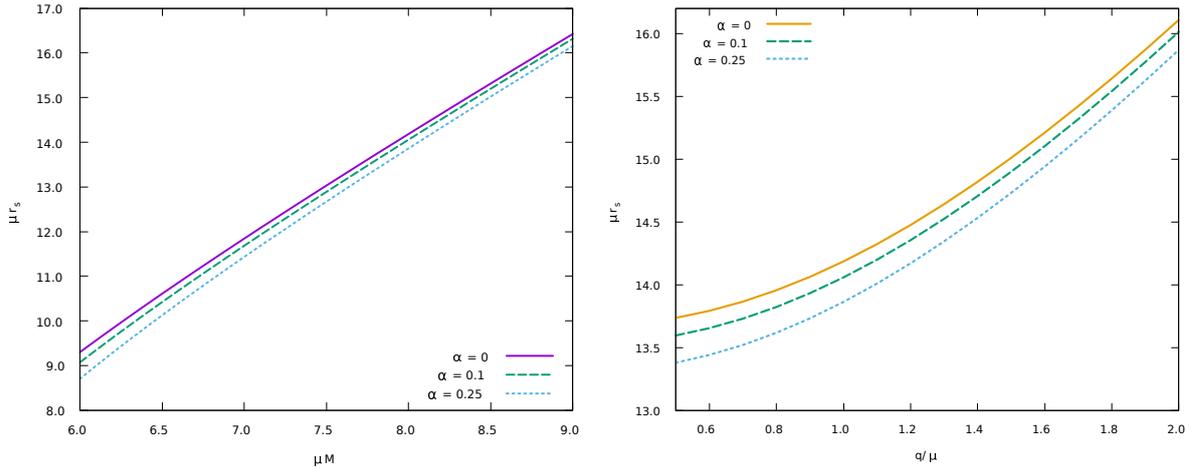}
\caption{Star radius dependence on the star mass with different {\it dark matter} coupling (left panel) and on the scalar field charge (right panel). In both plots $Q = 4,~Q_d = 4$, in the left panel $q = 1$ and in the right panel $M = 8$.}
\label{fig:rs_M_qf}
\end{figure}

The left panel of the figure \ref{fig:rs_M_qf} depicts the hairy star radius as a function of its mass.
Adding the {\it dark matter} coupling to the system shrinks the obtained radius.
A similar behavior is also shown on in the right panel of the figure \ref{fig:rs_M_qf}, where the radius versus scalar field charge is plotted.
Both of these plots are presented for equal charges $Q = Q_d = 4$.

%\begin{figure}
%\centering
%\includegraphics[width=0.6\textwidth]{rs_qf.eps}
%\caption{$Q = 4, Q_d = 4, M = 8$}
%\label{fig:rs_qf}
%\end{figure}

% USTALIĆ BEZWYMIAROWE JEDNOSTKI W RYSUNKACH!!!

%%%%%%%%%%%%%%%%%%%%%%
\section{Discussion and conclusions}
\label{sec3}
In our paper we have studied the charged spherically horizonless solution of Einstein {\it dark matter} sector equation of motion, being
the two $U(1)$-gauge theory in which the ordinary Maxwell field pertain to the {\it visible } sector while the auxiliary $U1)$-gauge field is responsible for the
{\it hidden} one. The solution in question models the ultra-compact star-like object in the underlying theory.

We have found the lower bound on the radius of the charged {\it dark matter} sphere, which takes place for any solutions satisfying the condition of pressure being less 
than the density of the object. However, it is possible to derive some refinement on the bound in question. Namely, the investigations of
photonsphere in the background of the spherically symmetric ultra-compact {\it dark matter} sector star objects reveals the fact that it has an upper bound expressed in terms of the ADM mass and total charge, which in turn gives us some additional condition binding the total charge and mass.

Due to the fact that spherically symmetric horizonless ultra-compact star-like object may be unstable to perturbations of massless fields, we obtain another restriction on its radius. It turns out that the larger {\it dark charge} is considered, the smaller radius of the compact object we get. When the charge and mass of the {\it dark matter} object
are constant, the increase of the $\alpha$-coupling constant value, binding the {\it hidden} and {\it visible} sectors, causes the decrease of the radius of {\it dark star}.

We have also investigated the possibility of condensation of scalar field around compact reflecting stars in a box-boundary regime.
The presented model includes the {\it dark matter} charge and its influence on the condensing scalar through the background metric.
From every plot presented in this article we can conclude that the {\it dark matter}--electromagnetic field coupling 
decreases the largest star radius which is capable of holding hair.
Moreover this effect is stronger when both charges are present in the system and when they are sufficiently large,
so the mixing term makes significant contribution in the metric function. On the other hand the effective charge may not be 
too large so the square root in \eqref{eq:rsrange} remains real.
We are aware of the restraints of the probe limit and 
that the problem requires a full study with back reaction and solving the Maxwell equations.
Also it is interesting to see the stability of the scalar condensate when the charged dark matter would be dropping into the system in a time-dependent approach.
We take it as an inspiration to further research to be presented elsewhere.


\begin{thebibliography}{99}

%%%%%%%%%%%%%%%%%%%%%%%%%%%%%%%%%%%%%%%%%%%%%%%%%%%%%%%%%%%%%%%%%%%%%%%%%%%%%%%%%%%%%%%%%
\def\cmp#1#2#3#4{\emph{#4}, \emph{ Commun. Math. Phys.} {\bf #1} (#3) #2}
\def\lmp#1#2#3#4{\emph{#4}, \emph{ Lett. Math. Phys.} {\bf #1} (#3) #2}
\def\hpa#1#2#3#4{\emph{#4}, \emph{ Hell. Phys. Acta} {\bf #1} (#3) #2}
\def\grg#1#2#3#4{\emph{#4}, \emph{ Gen. Rel. Grav.} {\bf #1} (#3) #2}
\def\pr#1#2#3#4{\emph{#4}, \emph{ Phys. Rev.} {\bf #1} (#3) #2}
\def\prl#1#2#3#4{\emph{#4}, \emph{ Phys. Rev. Lett.} {\bf #1} (#3) #2}
\def\prd#1#2#3#4{\emph{#4}, \emph{ Phys. Rev. D} {\bf #1} (#3) #2}
\def\prb#1#2#3#4{\emph{#4}, \emph{ Phys. Rev. B} {\bf #1} (#3) #2}
\def\pl#1#2#3#4{\emph{#4}, \emph{ Phys. Lett.} {\bf #1} (#3) #2}
\def\pla#1#2#3#4{\emph{#4}, \emph{ Phys. Lett. A} {\bf #1} (#3) #2}
\def\plb#1#2#3#4{\emph{#4}, \emph{ Phys. Lett. B} {\bf #1} (#3) #2}
\def\prep#1#2#3#4{\emph{#4}, \emph{ Phys. Reports} {\bf #1} (#3) #2}
\def\phys#1#2#3#4{\emph{#4}, \emph{ Physica} {\bf #1} (#3) #2}
\def\jcp#1#2#3#4{\emph{#4}, \emph{ J. Comput. Phys.} {\bf #1} (#3) #2}
\def\jmp#1#2#3#4{\emph{#4}, \emph{ J. Math. Phys.} {\bf #1} (#3) #2}
\def\jpm#1#2#3#4{\emph{#4}, \emph{ J. Phys. A: Math. Gen.} {\bf #1} (#3) #2}
\def\cpr#1#2#3#4{\emph{#4}, \emph{ Computer Phys. Rept.} {\bf #1} (#3) #2}
\def\cqg#1#2#3#4{\emph{#4}, \emph{ Class. Quant. Grav.} {\bf #1} (#3) #2}
\def\cma#1#2#3#4{\emph{#4}, \emph{ Computers Math. Applic.} {\bf #1} (#3) #2}
\def\mc#1#2#3#4{\emph{#4}, \emph{ Math. Compt.} {\bf #1} (#3) #2}
\def\apj#1#2#3#4{\emph{#4}, \emph{ Astrophys. J.} {\bf #1} (#3) #2}
\def\apjs#1#2#3#4{\emph{#4}, \emph{ Astrophys. J. Suppl.} {\bf #1} (#3) #2}
\def\acta#1#2#3#4{\emph{#4}, \emph{ Acta Astronomica} {\bf #1} (#3) #2}
%%%%%%%%%%%%%%%%%%%%%%%%%%%%%%%%%%%%%%%%%%%%%%%%%%%%%%%%%%%%%%%%%%%%%%%%%%
\def\apl#1#2#3#4{\emph{#4}, \emph{ Ann. Physik. (Leipzig)} {\bf #1} (#3) #2}
\def\amjp#1#2#3#4{\emph{#4}, \emph{Am. J. Phys.} {\bf #1} (#3) #2}
\def\anp#1#2#3#4{\emph{#4}, \emph{ Ann. Phys.} {\bf #1} (#3) #2}
\def\sa#1#2#3#4{\emph{#4}, \emph{ Sov. Astro.} {\bf #1} (#3) #2}
\def\sia#1#2#3#4{\emph{#4}, \emph{ SIAM J. Sci. Statist. Comput.} {\bf #1} (#3) #2}
\def\aa#1#2#3#4{\emph{#4}, \emph{ Astron. Astrophys.} {\bf #1} (#3) #2}
\def\mnras#1#2#3#4{\emph{#4}, \emph{ Mon. Not. R. Astr. Soc.} {\bf #1} (#3) #2}
\def\npb#1#2#3#4{\emph{#4}, \emph{ Nucl. Phys. B} {\bf #1} (#3) #2}
\def\prsla#1#2#3#4{\emph{#4}, \emph{ Proc. R. Soc. London, Ser. A} {\bf #1} (#3) #2}
\def\jhep#1#2#3#4{\emph{#4}, \emph{ JHEP} {\bf #1} (#2) #3}
\def\nuc#1#2#3#4{\emph{#4}, \emph{ Nuovo Cimento B } {\bf #1} (#3) #2}
\def\ijmp#1#2#3#4{\emph{#4}, \emph{ Int. J. Mod. Phys. D} {\bf #1} (#3) #2}
\def\atmp#1#2#3#4{\emph{#4}, \emph{ Adv. Theor. Math. Phys.} {\bf #1} (#3) #2}
\def\ptps#1#2#3#4{\emph{#4}, \emph{ Prog. Theor. Phys. Suppl.} {\bf #1} (#3) #2}
\def\ptp#1#2#3#4{\emph{#4}, \emph{ Prog. Theor. Phys.} {\bf #1} (#3) #2}
\def\lmp#1#2#3#4{\emph{#4}, \emph{ Lett. Math. Phys.} {\bf #1} (#3) #2}
\def\cpam#1#2#3#4{\emph{#4}, \emph{ Comm. Pure Appl. Math.}  {\bf #1} (#3) #2}
\def\adv#1#2#3#4{\emph{#4}, \emph{ Adv. Phys.}  {\bf #1} (#3) #2}
\def\zh#1#2#3#4{\emph{#4}, \emph{ Zh. Eksp. Teor. Fiz.}  {\bf #1} (#3) #2}
\def\mplb#1#2#3#4{\emph{#4}, \emph{ Mod. Phys. Lett. B} {\bf #1} (#3) #2}
\def\jams#1#2#3#4{\emph{#4}, \emph{ J. Austral. Math. Soc. B} {\bf #1} (#3) #2}
\def\appa#1#2#3#4{\emph{#4}, \emph{ Acta Phys. Polonica A} {\bf #1}, (#3) #2}
\def\nat#1#2#3#4{\emph{#4}, \emph{Nature} {\bf #1}, (#3) #2}
\def\science#1#2#3#4{\emph{#4}, \emph{Science} {\bf #1}, (#3) #2}
\def\arcmp#1#2#3#4{\emph{#4}, \emph{Annual Rev. of Cond. Matter Physics} {\bf #1}, (#3) #2}
\def\zphys#1#2#3#4{\emph{#4}, \emph{Z. Phys.} {\bf #1}, (#3) #2}
\def\ncs#1#2#3#4{\emph{#4}, \emph{Nuovo Cimento Suppl.} {\bf #1}, (#3) #2}
%
\def\hepph#1#2{{ hep-ph }{#1} (#2)}
\def\hepth#1#2{{ hep-th }{#1} (#2)}
\def\grqc#1#2{{ gr-qc }{#1} (#2)}
\def\ibid#1#2#3#4{\emph{#4}, {\it ibid.} {\bf #1} (#3) #2}
\def\conphy#1#2#3#4{\emph{#4}, \emph{Contemporary Physics} {\bf #1}, (#3) #2}
\def\jcap#1#2#3#4{\emph{#4}, \emph{JCAP} {\bf #1}, (#3) #2}
\def\rpp#1#2#3#4{\emph{#4}, \emph{Rep. Prog. Phys.} {\bf #1}, (#3) #2}
\def\njp#1#2#3#4{\emph{#4}, \emph{New J. Phys.} {\bf #1}, (#3) #2}
\def\jpg#1#2#3#4{\emph{#4}, \emph{ J. Phys. G} {\bf #1} (#3) #2}
\def\eurphysjc#1#2#3#4{\emph{#4}, \emph{ Eur. Phys. J.  C} {\bf #1} (#3) #2}
\def\eurphysjplus#1#2#3#4{\emph{#4}, \emph{ Eur. Phys. J.  Plus} {\bf #1} (#3) #2}

%%%%%%%%%%%%%%%%%%%%%%%%%%%%%%%%%%%%%%%%%%%%%%%%%%%%%%%%%%%%%%%%%%%%%%%%%%%%%%%
%%%%%%%%%%%%%%%%%%%%%%%%%%%%%%%%%%%%%%%%%%%%%%%%%%%%
\bibitem{ber18}
G. Bertone and T.M.P. Tait, \nat{562}{51}{2018}{A new era in the search for dark matter}.

%%%%%%%%%%%Scaffolding dm %%%%%%%%%%%%%%
\bibitem{mas07}
R. Massey et al., \nat{445}{286}{2007}{Dark matter maps reveal cosmic scaffolding}.
\bibitem{die12}
J. Dietrich et al., \nat{487}{202}{2012}{A filament of dark matter between two cluster of galaxies}.

%%%%%%%%%%%%%%%%%%%%%%%%%%%%%%%%%
%%%%%%%%%%%%%%%%%%%%%%%%%%%%%%%%%%%%%%%%%%%%%%%%%%%%%%%%%%%%%%%%%%%%%%%%%%%%%%%%%%%%
\bibitem{sta16}
T.K. Starkenburg, A. Helmi and L.V. Sales, \aa{587}{A24}{2016}{Dark influences}.

%%%%%%%%%%%%%%%%%%%%%%%%%%%%%%%%%%%%%%%%%%%%%%%%%%%%%%%%%%%%%%%%%%%%%
\bibitem{reg15}
M. Regis, J.Q. Xia, A. Cuoso, E. Branchini, N. Fornengo, and M. Viel, \prl{114}{241301}{2015}{Particle Dark Matter Searches Outside the Local Group}.
\bibitem{ali15}
Y. Ali-Haimoud, J. Chluba, and M. Kamionkowski, \prl{115}{071304}{2015}{Constrants on Dark Matter Interactions with Standard Model Particles from Cosmic Microwave Background Spectral Distortions}.
%%%%%%%%%%%%%%%%%%%%%%%%%%%%%%%%%%%%%%%%%%%%%%%%%%%%%%%%%%%%%%%%%%%%%%%%%%%%%%%%%%%%%%%%%%

%%%%%%%%%%%%%%%%%%%%%%%%%%DAMA %%%%%%%%%%%%%%
\bibitem{ber98}
R. Bernabei {\it et al.}, \plb{424}{195}{1998}{Searching for WIMPs by the annual modulation signature}.
\bibitem{ber13}
R. Bernabei {\it et al.}, \eurphysjc{73}{2648}{2013}{Final model independent result of DAMA/LIBRA-phase1}.


\bibitem{cos18}
The COSINE-100 Collaboration, \nat{564}{83}{2018}{An experiment to search for dark-matter interactions using sodium iodide detectors}.
\bibitem{masi2015}
N. Masi, \eurphysjplus{130}{69}{2015}{Dark matter: TeV-ish rather than miraculous, collision-less rather than dark}.



\bibitem{ger15}
A. Geringer-Sameth and M.G. Walker, \prl{115}{081101}{2015}{Indication of Gamma-Ray Emission from the Newly Discovered Dwarf Galaxy Reticulum II}.
\bibitem{bod15}
K.K. Boddy and J. Kumar, \prd{92}{023533}{2015}{Indirect detection of {\it dark matter} using MeV-range gamma-rays telescopes}.
\bibitem{til15}
K.Van Tilburg, N. Leefer, L. Bougas, and D. Budker, \prl{115}{011802}{2015}{Search for Ultralight Scalar Dark Matter with Atomic Spectroscopy}.
\bibitem{babar14}
J.P. Lees et al., \prl{113}{201801}{2014}{Search for a Dark Photon in $e^+ e^-$ Collisions at BABAR}.
%%%%%%%%%%%%%%%%%%%%%%%%%%%%%%%%%%%%%%%%%%%%%%%%%%%%%%%%%%%%%%%%%
\bibitem{cha17}
J.H. Chang, R. Essig, and S.D. McDermott, \jhep{01}{2017}{107}{Revisiting Supernova 1987A constraints on dark photons}.
\bibitem{massey15a}
D. Harvey, R. Massey, T. Kitching, A. Taylor and E. Tittley, \science{347}{1462}{2015}{The nongravitational interactions of {\it dark matter} in colliding galaxy clusters}.



%%%%%%%neutron stars, bh dark matter %%%%%%%%%%%%%%%%%%%%%%%%%%%%%%%%%%%%%%%%%%%%%%%%%%%%%%%
\bibitem{bra14}
J. Bramante and T. Linden, \prl{113}{191301}{2014}{Detecting {\it dark matter} with imploding pulsars in the galactic center}.
\bibitem{ful15}
J. Fuller and C.D. Ott, \mnras{450}{L71}{2015}{Dark-matter-induced collapse of neutron stars: a possible link between fast radio bursts and missing pulsar problem}.
\bibitem{lop14}
I. Lopes and J. Silk, \apj{786}{25}{2014}{A particle {\it dark matter} footprint on the first generation of stars}.
\bibitem{nak12}
A. Nakonieczna, M. Rogatko, and R. Moderski, \prd{86}{044043}{2012}{Dynamical collapse of charged scalar field in phantom gravity}.
\bibitem{nak15a}
A. Nakonieczna, M. Rogatko, and L. Nakonieczny, \jhep{11}{2015}{012}{\it Dark matter impact on gravitational collapse of an electrically charged scalar field}.



%%%%%%%%%%%%%%%%%%%% {\it dark matter} %%%%%%%%%%%%%%%%%%%%%%%%%%%%%%%%%%%%%%%%%%%%%%%%%%%
\bibitem{vac91}
T. Vachaspati and A. Achucarro, \prd{44}{3067}{1991}{Semilocal cosmic strings}.
\bibitem{ach00}
A. Achucarro and T. Vachaspati, \prep{327}{347}{2000}{Semilocal and electroweak strings}.


%%%%%%%%%%%%%%%%%alpha%%%%%%%%ALPHA %%%%%%%%%%%%%%%%%%%%%%%%%%%%%
\bibitem{abe04}
S.A. Abel and B.W. Schofield, \npb{685}{150}{2004}{Brane-antibrane kinetic mixing, millicharged particles and SUSY breaking}.
\bibitem{abe08}
S.A. Abel, J. Jaeckel, V.V.  Khoze, and A. Ringwald, \plb{666}{66}{2008}{Illuminating the hidden sector of string theory by shining light through a magnetic field}.
\bibitem{abe08a}
S.A. Abel, M.D. Goodsell, J. Jaeckel, V.V. Khoze, and A. Ringwald, \jhep{07}{2008}{124}{Kinetic mixing of the photon with hidden U(1)s in string phenomenology}.
\bibitem{ban17}
D. Banerjee et al., \prl{118}{011802}{2017}{Search for invisible decays of sub-GeV dark photons in missing-energy events at the CERN SPS}.





%%%%%%%%%%%%%%%%%%% {\it dark matter} model%%%%%%%%%%%%%%%%%%%%%%%%%%%%%%%%%%%%%%%%%%%
\bibitem{integral}
P. Jean {\it et al.}, \aa{407}{L55}{2003}{Early SPI/INTEGRAL measurements of 511 keV line emission from the 4th quadrant of the Galaxy}.
\bibitem{atic}
J. Chang {\it et al.}, \nat{456}{362}{2008}{An excess of cosmic ray electrons at energies of 300-800 GeV}.
\bibitem{pamela}
O. Adriani {\it et al.} (PAMELA Collaboration), \nat{458}{607}{2009}{An anomalous positron abundance in cosmic rays with energies 1.5-100 Gev}.
\bibitem{muon}
G.W. Bennett {\it et al.}, \prd{73}{072003}{2006}{Final report of the E821 muon anomalous magnetic moment measurement at BNL}.

\bibitem{dav12}
H. Davoudiasl, H.S. Lee and W.J. Marciano, \prd{85}{115019}{2012}{"Dark" Z implications for parity violation, rare meson decays, and Higgs physics}.
\bibitem{dav13}
H. Davoudiasl, H.S. Lee, I.Lewis and W.J. Marciano, \prd{88}{015022}{2013}{Higgs decays as a window into the dark sector}.



%
%%%%%%%%%%%%%%%%%%%%%%%%%%%%%%%%%%%%%%%%%%%%%%%%%%%%%%%%%%%%%%%%%%%%%%%%%%
%%%%%%%%%%%%%%%%%%%%%%%%%%%%%%    TOP-DOWN%%%%%%%%%%%%%%%%%%%%%%%%%%%%%%%
\bibitem{ach16}
B.S. Acharya, S.A.R. Ellis, G.L. Kane, B.D. Nelson and M.J. Perry,\prl{117}{181802}{2016}{ Lightest Visible-Sector Supersymmetric Particle is Likely Unstable}.
\bibitem{hol86}
B. Holdom, \plb{166}{196}{1986}{Two U(1)'s and $\epsilon$ charge shifts}.
\bibitem{portal1}
D. L\"ust, \cqg{21}{S1399}{2004}{Intersecting brane worlds: a path to the standard model?}.
\bibitem{portal2}
S. Abel and J. Santiago, \jpg{30}{R83}{2004}{Constraining the string scale: from Planck to weak and back again}.
\bibitem{abe08}
S.A. Abel, M.D. Goodsell, J. Jaceckel, V.V. Khoze, and A. Ringwald, \jhep{07}{2008}{124}{Kinetic mixing term of photon with hidden U(1)s in string phenomenology}.
%\bibitem{abe04}
%S.A.Abel and B.W.Schofield, \npb{685}{150}{2004}{Brane-antibrane kinetic mixing term, millicharged particles and SUSY braeking}.
%\bibitem{die97}
%K.R.Dienes, C.F.Kolda, J.March-Russel, \npb{492}{104}{1997}{Kinetic mixing and the supersymmetric gauge hierarchy}.

%%%%%%%%%%%%%%%%%%%%%%%%%%%%%%%%%%%%


%%%%%%%%%%%%%%%%%%%%%%%%%%%%%%%%%%%%%%%%%%%%%%%%%%%%%%%%%%%%%%%%%%%%%%%%%
\bibitem{kar08}
P. Karageorgis and J. Stalker, \cqg{25}{195021}{2008}{Sharp bounds on 2m/r for static spherical objects}.
\bibitem{and09}
H. Andreasson, \cmp{288}{715}{2009}{Sharp bounds on the critical stability radius for relativistic charged spheres}.
\bibitem{hod18}
S. Hod, \prd{98}{064014}{2018}{Upper bound on the gravitational masses of stable spatially regular charged compact objects}.

\bibitem{car09}
V. Cardoso, A.S. Miranda, E. Berti, and V.T. Zanchin, \prd{79}{064016}{2009}{Geodesic stability, Lyapunov exponents, and quasi-normal modes}.
\bibitem{nar05}
R. Narayan, \njp{7}{199}{2005}{Black holes in astrophysics}.

\bibitem{pen19}
Y. Peng, \plb{790}{396}{2019}{Upper bound on the radii of regular ultra-compact star photonspheres}.

%%%%%%%%%%%%%%%%%%%%%%%%light rings %%%%%%%%%%%%%%%%%%%%%%%
\bibitem{car14}
V. Cardoso, L.C.B. Crispino, C.F.B. Macedo, H. Okawa, and P. Pani, \prd{90}{044069}{2014}{Light rings as observational evidence for event horizons: 
Long-lived modes, ergoregions and nonlinear instabilities of ultracompact objects}.
\bibitem{cun17}
P.V. Cunha, E. Berti, and C.A.R. Herdeiro, \prl{119}{251102}{2017}{Light-ring stability for ultracompact objects}.
\bibitem{hod18pl}
S. Hod, \plb{776}{1}{2018}{On the number of light rings in curved spacetimes of ultra-compact objects}.
\bibitem{kei16}
J. Keir, \cqg{33}{135009}{2016}{Slowly decaying waves on spherically symmetric spacetimes and ultracompact neutron stars}.






%%%%%%%%%%%%%%%%%%%hair %%%%%%%%%%%%%%%%%%%%%%%%%%%%%%%%%%%%%%%%%%%%%%%
\bibitem{hod16}
S. Hod, \prd{94}{104073}{2016}{No-scalar hair theorem for spherically symmetric reflecting stars}.
\bibitem{bha17}
S. Bhattacharjee, \prd{95}{084027}{2017}{No-hair theorem for static and stationary reflecting star}.
\bibitem{hod17}
S. Hod, \prd{96}{024019}{2017}{No nonminimally coupled scalar hair for spherically symmetric neutral reflecting stars}.
\bibitem{hod16b}
S. Hod, \plb{763}{275}{2016}{Charged massive scaalr field configurations supported by a spherically symmetric charged reflecting shell}.
\bibitem{hod17b}
S. Hod, \plb{768}{97}{2017}{Marginally bound resonances of charged massive scalar fields in the background of a charged reflecting shell}.
\bibitem{her13}
C.A. R. Hedeiro, J.C. Degollado, and H.F. Runarsson, \prd{88}{063003}{2013}{Rapid growth of superradiant instabilities for charged black holes in a cavity}.
\bibitem{san16}
N. Sanchis-Gual, J.C. Degollado, P.J. Montero, and C. Hedeiro, \prl{116}{141101}{2016}{Explosion and final state of an unstable Reissner-Nordstrom black hole}.
\bibitem{pen18}
Y. Peng, \plb{780}{144}{2018}{Scalar field configurations supported by charged compact reflecting stars in a curved spacetime}.

















\end{thebibliography}
\end{document}